# Room-temperature ferromagnetism in graphite driven by 2D networks of point defects

J. Červenka[1]†, M. I. Katsnelson[2], and C. F. J. Flipse[1]*

[1] *Department of Applied Physics, Eindhoven University of Technology, 5600 MB Eindhoven, The Netherlands*

[2] *Institute of Molecules and Materials, Radboud University of Nijmegen, NL-6525 ED Nijmegen, The Netherlands*

† Present address: *Institute of Physics, Academy of Sciences of the Czech Republic, Cukrovarnická 10, CZ-162 53 Prague 6, Czech Republic*

* e-mail: C.F.J.Flipse@tue.nl

**Ferromagnetism in carbon-based materials is appealing for both applications and fundamental science purposes because carbon is a light and bio-compatible material that contains only** *s* **and** *p* **electrons in contrast to traditional ferromagnets based on 3***d* **or 4***f* **electrons. Here we demonstrate direct evidence for ferromagnetic order locally at defect structures in highly oriented pyrolytic graphite (HOPG) with magnetic force microscopy and in bulk magnetization measurements at room temperature. Magnetic impurities have been excluded as the origin of the magnetic signal after careful analysis supporting an intrinsic magnetic behavior of carbon. The observed ferromagnetism has been attributed to originate from unpaired electron spins localized at grain boundaries of HOPG. Grain boundaries form two-dimensional arrays of point defects, where their spacing depends on the mutual orientation of two grains. Depending on the distance between**

**these point defects, scanning tunneling spectroscopy of grain boundaries showed two intense split localized states for small distances between defects (< 4 nm) and one localized state at the Fermi level for large distances between defects (> 4 nm).**

Ferromagnetism in carbon-based materials is controversial since only *sp* electrons are present, magnetic signals are very small and the Curie temperature exceeds room temperature. However, several independent observations have been reported to confirm the existence of the ferromagnetic order in impurity-free carbon materials[1-4]. The ferromagnetism in graphitic materials showed to be closely related to the lattice imperfections as demonstrated by induced ferromagnetism in proton-irradiated graphite spots[2] or by increased magnetic signals in specially prepared pyrolytic graphite containing a high defect concentration[3]. Beside graphite, ferromagnetism has been observed in other carbon-based materials such as polymerized fullerenes[7], carbon nanofoam[8], proton irradiated thin carbon films[9], and nitrogen and carbon ion implanted nanodiamond[10]. All these observations suggest an inherent ferromagnetic behavior of carbon-based materials.

Several theoretical investigations have been carried out to explain magnetism observed in these systems. The origin of ferromagnetism was suggested to be attributed to the mixture of carbon atoms with alternating $sp^2$ and $sp^3$ bonds[11], the presence of a negatively curved graphitic surface containing seven- or eight-membered rings[12], and the existence of zigzag edges[13-15]. Recently, it has been shown in spin-polarized density functional theory (DFT) calculations that point defects in graphite such as vacancies and hydrogen-terminated vacancies are magnetic[17,18]. Randomly distributed single-atom defects have demonstrated ferromagnetism in disordered graphite with preserved stacking order of graphene layers[19]. Three-dimensional network of single-atom vacancies in graphite developed ferrimagnetic ordering up to 1 nm separation among the vacancies[20].

Although ferromagnetic signals have been detected in graphite before[1-6], the origin of the ferromagnetism remained unknown. Here we report an experimental observation of ferromagnetic order in HOPG detected specifically at defect structures. A ferromagnetic signal has been observed locally with magnetic force microscopy (MFM) and in the bulk magnetization measurements using superconducting quantum interference device (SQUID). A theoretical model is introduced to qualitatively explain the MFM and SQUID observations on the base of 2D periodical network of point defects at grain boundaries of HOPG.

Atomic force microscopy (AFM), magnetic force microscopy (MFM) and electrostatic force microscopy (EFM) images of the same area on the HOPG surface are shown in Fig. 1. The AFM topography picture in Fig. 1a displays a surface with a high population of step edges, surface distortions and defects. The MFM images in Figs 1b and 1c were taken on the same place as the AFM image with a lift scan height of 50 nm, where long-range van der Waals forces are negligible and magnetic forces prevail. A magnetic signal is measured on most of the line defects, while a step edge marked as A in Fig. 1a does not show a magnetic signal in the MFM image. On the other hand, two lines in the MFM image in Fig. 1b that are indicated as B and C do not show a noticeable height difference in the topography. The lines B and C are grain boundaries of HOPG. Their detailed AFM and STM study can be found elsewhere[21,22].

In order to determine the character of the detected magnetic signal, the MFM tip has been magnetized in two opposite directions: pointing into (Fig. 1b) and out of the graphite surface plane (Fig. 1c). Since the MFM signal represents the phase shift between the probe oscillation and the driving signal due to magnetic force acting on the tip, the dependence of the phase shift on the force gradient can be expressed by a simple form[23]

$$\Delta\Phi \approx \frac{Q}{k}\frac{\partial \mathbf{F}}{\partial z},\qquad(1)$$

where $Q$ is quality factor and $k$ is spring constant of the cantilever. Typical values of our MFM system give a minimal detectable force gradient in the order of 100 μN/m, $Q = 200$ and $k = 2.8$ N/m. For a true quantitative interpretation of MFM images it is necessary to have an exact knowledge of the geometry and magnetic properties of the tip and the substrate in order to express the force acting on the tip, which is difficult and has been achieved only in special cases[23]. Nevertheless, a qualitative analysis can be done according to expression 1, where a positive phase shift (bright contrast) represents a repulsive force between the tip and the sample, and a negative phase shift (dark contrast) manifests an attractive interaction relative to the background signal. Since the tip magnetized into the graphite surface plane has shown a bright contrast in Fig. 1b and out of plane magnetized tip produced a dark phase contrast on the line defects in Fig. 1b, the orientation of the net magnetic moment in the defects stayed in the same direction, pointing out of the graphite surface plane. This shows a clear indication of ferromagnetic order at the defect sites at room temperature. In the case of paramagnetic order, a bright contrast would be detected in both direction of the magnetization of the tip because the local magnetic moments would align with the magnetic field of the tip leading to attractive interaction. The same result would be valid if electric force gradients were detected due to charge accumulation at the step edges. Therefore, the ferromagnetic order in the defects of the HOPG sample is the only plausible explanation for the detected MFM signal.

However, not all the signal measured in the MFM showed to be sensitive to the reversal of the tip magnetization, in particular, areas with a different phase contrast. This is due to the metallic character of the magnetic coating film of the MFM tip, which probes electrostatic forces as well. Therefore EFM has been measured on the same place with Pt coated Si tip with a lift scan height of 20 nm (see Fig. 1d). A bright contrast is observed on the same places as in the MFM images. Similar observations of regions with a different potential has been measured in EFM and Kelvin probe microscopy (KPM) on HOPG before[24,25]. This non-uniform potential distribution has been found to be caused by the mechanical stress induced during sample cleaving[25]. Thereby, the MFM measurements represent a superposition of magnetic and electrostatic signal, which explains well the observed line shapes in Fig. 1c.

The magnetization analysis of the HOPG samples has been performed with a SQUID magnetometer at 5 K and 300 K. Figures 2 show out-of-plane (along c-axis) and in-plane magnetization (perpendicular to *c*-axis) loops of HOPG after subtraction of linear diamagnetic background signals. Ferromagnetic-like hysteresis loops are observed both at 5 K and 300 K. The saturation magnetization reaches the largest value 0.013 emu/g in the out-of-plane orientation at 5 K. The in-plane magnetization loops are comparable to previous SQUID measurements on HOPG reported by P. Esquinazi *et al.*[3]. The in-plane magnetization loops saturate at a 5 times smaller value than in the out-of-plane configuration at 5 K. The coercive field and remnant magnetization are similar in both in-plane and out-of-plane magnetization measurements. In the work of P. Esquinazi *et al.*[3], the ferromagnetic signals were measured up to temperature 500 K.

The observed high temperature ferromagnetism in HOPG can have different possible origins. The first one is obviously ferromagnetism due to magnetic impurities. HOPG samples, as it has been studied previously[2-4], contain small fraction of magnetic elements. Therefore, we have analyzed the HOPG samples for impurity concentration by particle induced X-ray emission (PIXE) in the bulk material and by low energy ion scattering (LEIS) at the surface. As a main magnetic impurity in PIXE was found Fe with concentration ≈ 20 µg/g. Other magnetic and metallic impurities have been found below 1 µg/g. The surface analysis by LEIS has not detected any magnetic elements indicating impurity concentration below 100 ppm. The measured content of Fe impurities in HOPG is not sufficient to produce the ferromagnetic signal shown in Fig. 2. The amount of 1 µg/g of Fe would contribute maximally $2.2 \times 10^{-4}$ emu/g to the magnetization and for Fe or $Fe_3O_4$ clusters, the magnetic signals would be even smaller[3].

Another possible source of the shown up ferromagnetic behavior are the defect structures in graphite. Line defects occur naturally in graphite as edges and grain boundaries. Graphite edges have been extensively studied both theoretically[13-16] and experimentally[26-28]. There are two typical shapes for graphite edges: armchair and zigzag. Only zigzag edges are expected to give rise to the magnetic ordering due to the existence of the edge state[13]. STM experimental results on step edges of graphite, however, showed that zigzag edges are much smaller in length (≈ 2 nm) than those of armchair edges and less frequently observed[28]. Moreover, due to the one-dimensional character zigzag edges are not expected to maintain the ferromagnetic order at room temperature. The long-range magnetic order at the zigzag edges was predicted to be ~1 nm at 300 K.[15] Hence graphite edges could not produce the magnetic signals in MFM at room temperature. We rather believe that some of step edges are created on HOPG surfaces at places where bulk grain boundaries cross the surface. During the cleavage of HOPG, grain boundaries are the weakest points of the graphite crystal. A step edge created in this way would have the same orientation and geometry as a grain boundary underneath.

Grain boundaries in HOPG have been studied in great detail by AFM and STM[21,22]. Grain boundaries are inevitable defects in graphite because of polycrystalline character of HOPG. They are formed between two grains during the crystal growth and therefore they extend over step edges and form a continuous network all over the graphite surface. Grain boundaries show a small or no apparent height in AFM (see Fig. 1). On the other hand, they exhibit a very distinct sign in STM, where they appear as one-dimensional superlattices with a height corrugation up to 1.5 nm due to a charge accumulation. Figure 3 shows STM images and STS spectra on two typical grain boundaries with different periodicities. STS on grain boundaries exhibits localized electron states that are not present on the bare graphite surface. Grain boundaries with small distances between their defects <4 nm are characterized by two split localized electron states, while grain boundaries with large distances between their defects >4 nm display only one localized state, similarly like solitary defects in graphite[29]. Two localized states of grain boundaries are located predominantly around -0.2 V and 0.4 V. Grain boundaries with two localized states have been observed on graphite surfaces more frequently (80%) as can be seen from the statistics in Fig. 3. Due to the localized states, grain boundaries could be of the origin of the observed ferromagnetism in HOPG.

Defects in graphene break the translational symmetry of the lattice, which leads to creation of localized states at the Fermi energy and to the phenomenon of self-doping, where charge is transferred to/from defects to the bulk[30,31]. The graphene lattice is a bipartite lattice, which is formed by two interpenetrating triangular sublattices of carbon atoms (labeled A and B), such that the first neighbors of an atom A belong to the sublattice B and vice versa. Lieb has proven using Hubbard model and neutral bipartite lattice that the total spin $S$ of the ground state is given by $2S = |N_A - N_B|$, where $N_A$ is a number of atoms in sublattice A and $N_B$ in sublattice B[32]. Thus, Lieb's theorem states that a sublattice unbalance causes always a finite total magnetic moment in the graphene lattice. This imbalance can be induced for instance by single-atom vacancies, which remove only one atom of the sublattice, or by multiple vacancies where more A or B atoms are removed. Since graphene systems have low electron densities at the Fermi energy, electron-electron interactions play an important role as the recent experiments showed[33]. In the presence of local electron-electron interaction the localized states will become polarized, leading to the formation of local moments[31]. This has been illustrated in DFT studies of point defects in graphite such as vacancies and hydrogen-terminated vacancies. These defects revealed to be magnetic having a local magnetic moment larger than $1\mu_B$.[17,18] In the DFT study of a 3D array of single vacancies in graphite, different supercells containing single-atom vacancies have been studied[20]. Ferrimagnetic order has been supported up to a distance of 1 nm among the vacancies, while $5\times5\times1$ supercell (1.23 nm separated vacancies) did not show a net magnetic moment in graphite[20]. Two spin-polarized localized states have been observed at the vacancy site for vacancy distances up to 1 nm, while only one localized peak at the Fermi energy was formed for larger separation between vacancies similarly like for an isolated vacancy. In graphene, the $5\times5$ supercell exhibited still a net magnetic moment of $1.72\mu_B$.[20]

In a similar way, grain boundaries in graphite can be visualized as a two-dimensional plane of equidistantly distributed defects (see Fig. 4), where the distance between defects is given by the superlattice periodicity in the graphene plane and by the graphene layer separation 0.335 nm. The defects in grain boundaries are not single vacancies, for which a simple trigonal symmetry would be expected to be observed in STM[34], but rather more complicated defects. In Fig. 4, two characteristic model structures of grain boundaries on the graphite surface are shown. The first structure of a grain boundary is characterized by periodicity $D = d/2\sin(\alpha/2)$, where $d$ is the graphite lattice periodicity and $\alpha$ is an angle between two graphite grains. The orientation of this grain boundary has direction slightly off the armchair edge by angle $30°-\alpha/2$. This results in creation of a periodic array of defects, where their atomic structure along the axis of a grain boundary consists of long armchair edges alternated by short zigzag edges. In this structure, a sublattice unbalance is created owing to existence of a zigzag segment within an armchair edge. One segment of zigzag edge removes similarly like single-atom vacancy one of the atom sublattices of sublattice A or B. Therefore, $N_A \neq N_B$ and the local magnetic moment is created in accordance with Lieb's theorem. The second characteristic structure of a grain boundary has $\sqrt{3}D$ periodicity and is rotated by 30° with respect to the previous structure. Hence the internal structure of such a grain boundary is characterized by long zigzag edges and by short armchair edges as shown in Fig. 4b. The structure of such a grain boundary is symmetric, therefore $N_A = N_B$, even though long zigzag edges are present. However, a local magnetic moment can be formed in analogy to the extended vacancy shown in ref. 35. Since the defects in both grain boundary structures are created by zigzag segments, which lie on the same sublattice, the coupling between the local movements in the grain boundary would be always ferromagnetic[19,35].

An estimate of magnetic moment per one defect of a grain boundary leads to 0.2-1.5 $\mu_B$ per defect, assuming distances between defects of grain boundaries 0.5-4 nm, $M_S$ = 0.013 emu/g ≈ $2.6 \times 10^{-5}$ $\mu_B$ per carbon atom and uniform concentration of line defects ≈ 500 ppm (determined from MFM). This magnetic moment is in accordance to the theoretically predicted value of magnetic moment 1.12-1.53 $\mu_B$ of one vacancy site in graphene[18]. If the spin polarized electron states were created in a grain boundary, the exchange splitting would be in order of 0.6 eV in our experiment (see Fig. 3b). Another supporting evidence that the ferromagnetism originates from grain boundaries is their two dimensional character, which can explain most of the features from MFM and SQUID measurements.

We assume that grain boundaries are propagating along the *c*-axis of the graphite crystal creating 2D plane of defects. As it was described before, step edges can be the manifestation of the grain boundaries buried underneath them. The ferromagnetic signal would then come from 2D grain boundary planes formed through the bulk crystal. Moreover, an infinitely extended 2D magnetic plane with in-plane magnetization is stray-field-free and therefore it can exist in the single-domain state[36]. Accordingly, in-plane magnetized grain boundary plane should show a single magnetic domain, supporting the observation of only one magnetization direction in MFM measurements (Fig. 1). Due to crossings among grain boundaries, the minimum energy configuration would lead to magnetization pointing along the *c*-axis of HOPG. Magnetic field gradients from the edge of 2D grain boundary decay as $dB/dz \approx 1/z^2$, which gives rise to an estimation of force gradients $10^{-3}$-$10^{-4}$ N/m at 50 nm lift height in MFM, by using analogy to MFM simulations of Fe nanoparticles[37]. On the other hand, solitary Fe nanoparticles with a core size 10 nm would not be detectable in MFM at 50 nm lift scan height because of the fast decay of their magnetic field gradients $1/z^4$, resulting in 2500 times smaller magnetic field gradients than from 2D magnetic planes. 2D character of grain boundaries supports also the higher out-of-plane saturation magnetization of HOPG (parallel to the *c*-axis). The in-plane magnetization contribution of HOPG is measured because grain boundaries do not lie exactly in the *c*-axis but have a small tilt δ (see Fig. 4c). This angle of deviation of the grain's boundary from the perpendicular axis is given by the mosaic spread of HOPG, which is 3.5° - 5° for our samples. Therefore, a larger magnetic field is necessary to align the local magnetic moments of grain boundaries along the *c*-axis than along the graphene planes, where the magnetic axis stays in the 2D grain boundary plane. Hence magnetization measured in-plane of HOPG shows easy magnetic axis and out-of-plane magnetization demonstrates a hard magnetic axis (Fig. 2). Interestingly, the anisotropic signals found in SQUID measurements agree well with the spin resonance results in graphite of Wagoner[38]. The *g*-value of the resonance has shown remarkably large anisotropy with a strong temperature dependence[38]. While *g* remained temperature independent in the basal plane of graphite with $g_{IN}$ = 2.0026, it has grown from $g_{OUT}$ = 2.049 at 300 K to $g_{OUT}$ = 2.127 at 77 K in the direction parallel to the *c*-axis[38]. The g-value anisotropy has thus increased 2.5 times from room temperature to 77 K in the out-of plane direction [38], similarly like the saturation magnetization in Fig. 2. The origin of the *g*-value anisotropy in graphite, however, remains still unexplained even after 50 years of its first observation.

Ferromagnetic order in graphite demonstrate unexpectedly high Curie temperature reaching values well above room temperature (>500 K) as reported in other studies[3,5]. The temperature behavior of 2D grain boundary plane containing local magnetic moments can be described by the 2D anisotropic Heisenberg model[39]. Unlike 1D or 2D isotropic magnets which possess long-range order only in the ground state, real 1D and 2D magnets have shown finite values of the magnetic ordering temperature $T_C$ due to weak interlayer coupling and/or magnetic anisotropy[39]. The 2D anisotropic Heisenberg model using self-consistent spin-wave theories (SSWT) with Dyson-Maleev, Schwinger and combined boson-pseudofermion representations has been recently developed to describe magnetism in layered magnetic materials[39]. This model due to correct fluctuation corrections to SSWT has successfully described behavior of several layered magnets such as $La_2CuO_4$, $K_2NiF_4$ and $CrBr_3$ leading to an excellent agreement with experimental values of $T_C$. The analytical results for the Curie temperature was obtained

$$T_C = 4\pi JS^2 \left[ \ln \frac{T_C}{JS\Delta_0} + 4\ln \frac{4\pi JS^2 \Delta_0}{T_C} + C_F \right]^{-1}, \qquad (2)$$

where $J$ denotes an exchange integral, $S$ is a total spin of a defect, $\Delta_0$ is the dimensionless energy spin-wave gap, and constant $C_F$ gives only a small contribution (for details see ref. 39). If we use the results obtained by first principle calculations for zigzag graphene edges[15]: $S = 1/2$, $J = 4a = 420$ meV and $\Delta_0 = 10^{-4}$, the Curie temperature of 2D magnetic grain boundary would be $T_C = 764$ K. This result gives the low limit of the Curie temperature. If larger values of $\Delta_0$ or the total spin $S$ of a defect within grain boundary were used the critical temperature would be only larger. According to the above analysis, we believe that the grain boundaries are the most possible source of magnetism in graphite feasible to reach high Curie temperatures well above room temperature.

In conclusion, ferromagnetic signals have been observed in HOPG by magnetic force microscopy and SQUID magnetization measurements at room temperature. The observed ferromagnetism has been attributed to originate from unpaired electron spins localized at defects sites of grain boundaries. STM and STS have revealed localized states and enhanced charge density of grain boundaries. Grain boundaries in graphite, due to the special atomic structures, hold an important key for magnetic moment formation and for possible high temperature ferromagnetic order in graphitic materials.

**Acknowledgement:**

We are very grateful to R. Lavrijsen for SQUID measurements, P. H. A. Mutsaers for PIXE analysis and H. H. Brongersma for LEIS measurements. We thank B. Koopmans and H. J. M. Swagten for fruitful discussions and comments on the manuscript. This research was supported by Nanoned.

**Methods:**

Samples of HOPG of ZYH quality were purchased from NT-MDT. The ZYH quality of HOPG with the mosaic spread 3.5°-5° has been chosen because it provides a high population of step edges and grain boundaries on the graphite surface. HOPG samples were cleaved by an adhesive tape in air and transferred into a scanning tunneling microscope (Omicron LT STM) working under ultra high vacuum (UHV) condition. The HOPG samples have been heated to 500° in UHV before the STM experiments. STM measurements were performed at 78 K in the constant current mode with mechanically formed Pt/Ir tips. The same samples have been subsequently studied by AFM, MFM and EFM in air using Dimension 3100 SPM from Veeco Instruments. PPP-MFMR cantilevers made by NanoSensors and MESP cantilevers from Veeco Instruments with hard magnetic material Co-coating films have been used in the MFM tapping/lift mode.

**Figure captions:**

**Figure 1:** The same area on the HOPG surface imaged with AFM (**a**), MFM (**b**) and (**c**), and EFM (**d**). MFM tip has been magnetized into the graphite surface (**b**) and out of the graphite surface (**c**), respectively. Image parameters: scan area $2 \times 2$ μm$^2$, AFM $z$-range $z = 5$ nm, MFM $z$-range (**b**) $\Phi = 2°$ and (**c**) $\Phi = 1°$, the MFM lift height $h = 50$ nm, EFM $z$-range $\Phi = 1°$, the EFM lift height $h = 20$ nm.

**Figure 2:** Out-of plane (**a,b**) and in-plane (**c,d**) SQUID magnetization measurements on HOPG after subtraction of the diamagnetic signals at 5 K and 300 K. Magnetic field has been applied along the *c*-axis in the out-of-plane direction and along the graphene planes in the in-plane direction. The diamagnetic background signals were (a) $\chi_{OUT}$ = -1.1 × 10$^{-5}$ emu/g mT, (b) $\chi_{OUT}$ = -6.8 × 10$^{-6}$ emu/g mT, (c) $\chi_{IN}$ = -5.4 × 10$^{-7}$ emu/g mT, $\chi_{IN}$ = -3.9 × 10$^{-7}$ emu/g mT.

**Figure 3:** STM image of a grain boundary on HOPG showing a 1D superlattice with a small periodicity *D* = 1.4 nm (**a**) and a large periodicity *D* = 4 nm (**b**). Scanning parameters: (**a**) 10 × 10 nm$^2$, *U* = 0.6 V, *I* = 0.4 nA; and (**b**) 30 × 30 nm$^2$, *U* = 1 V, *I* = 0.06. (**c**) STS on two grain boundaries and on the bare graphite surface (tunneling resistance 0.9 GΩ). The grain boundary with *D* = 2.6 nm shows two localized states at -0.27 V and 0.4 V and the grain boundary with *D* = 4 nm demonstrates only one localized state at the Fermi level. (**d**) Energy positions of localized states measured on 15 different grain boundaries plotted against their superlattice periodicity.

**Figure 4:** Models of two basics shapes of grain boundaries in graphite: (**a**) armchair direction with periaodicity *D* and (**b**) zigzag direction with periodicity $\sqrt{3}D$. (**c**) 2D in-plane magnetized grain boundary propagating through bulk HOPG.

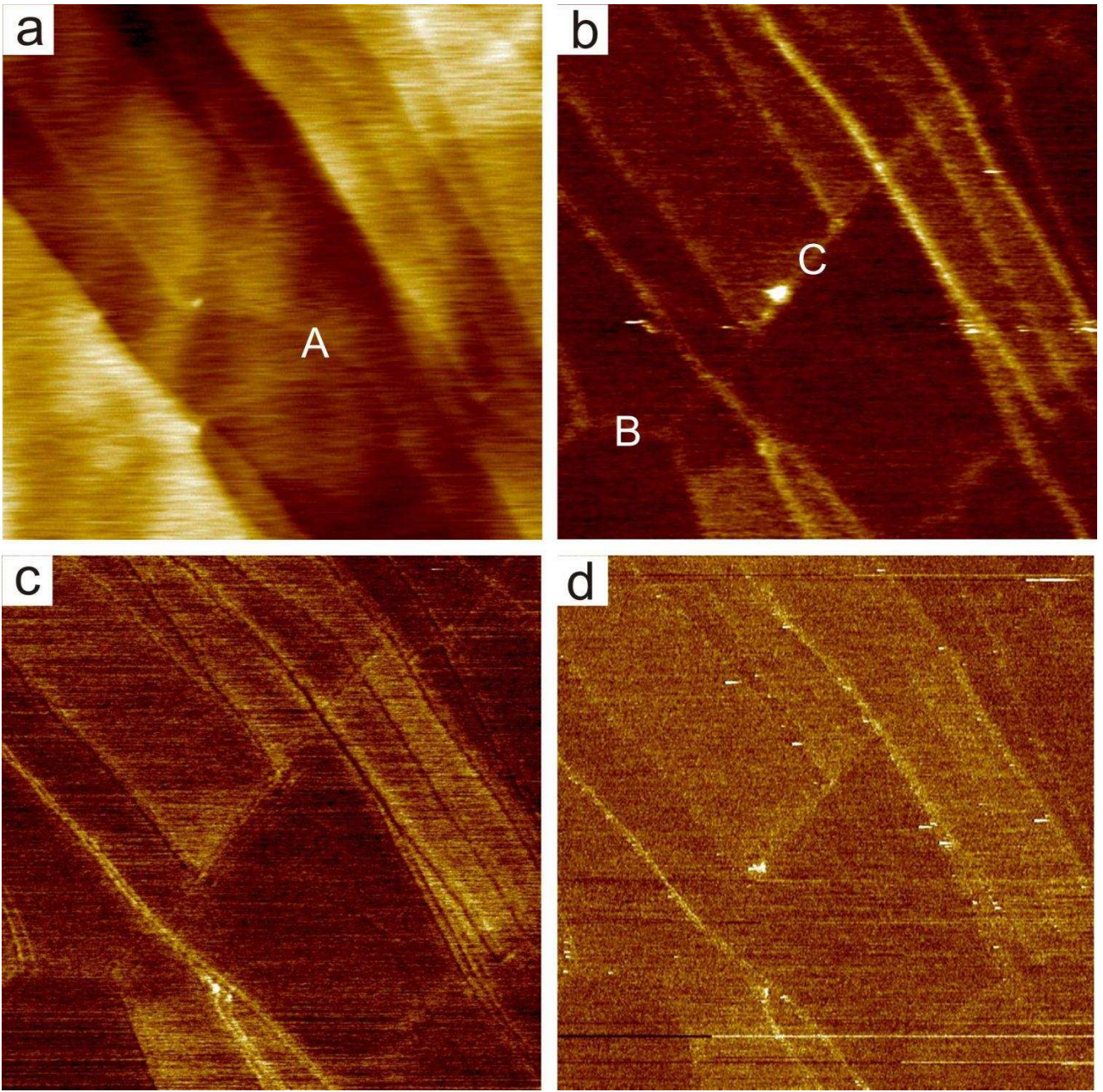

**Figure 1**

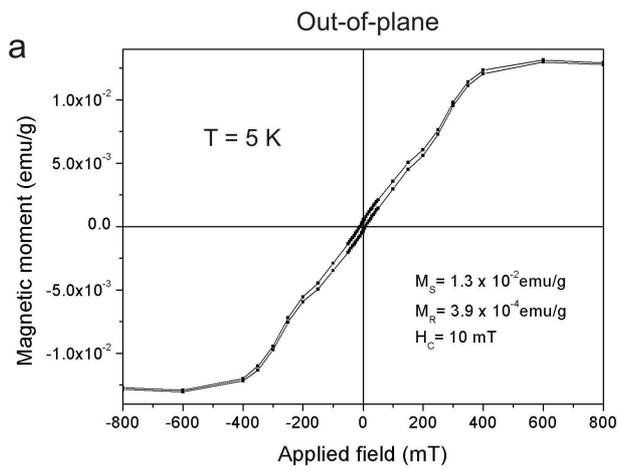
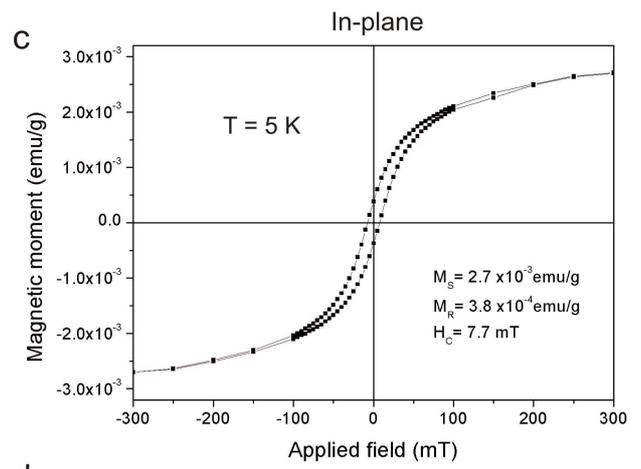
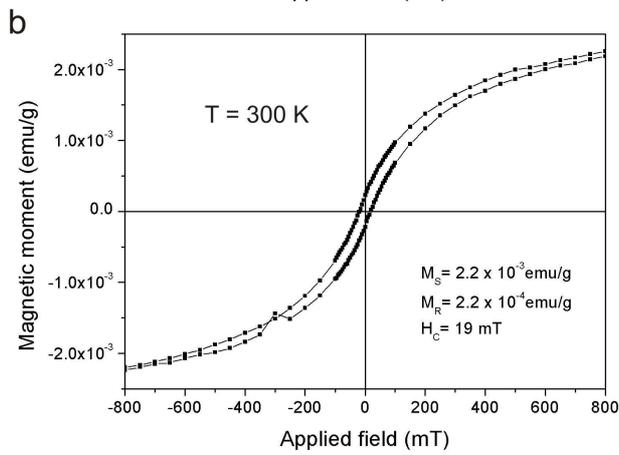
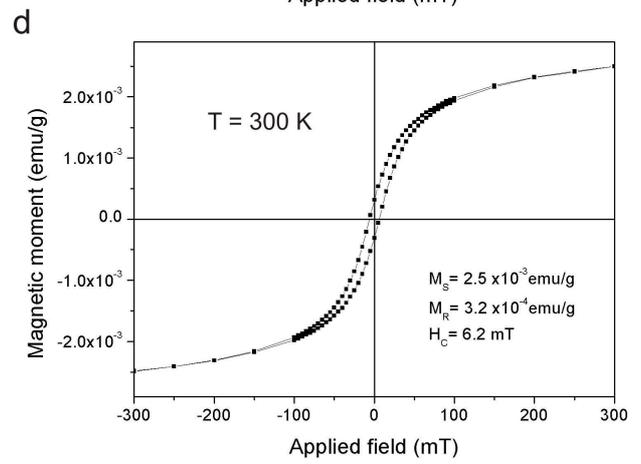

**Figure 2**

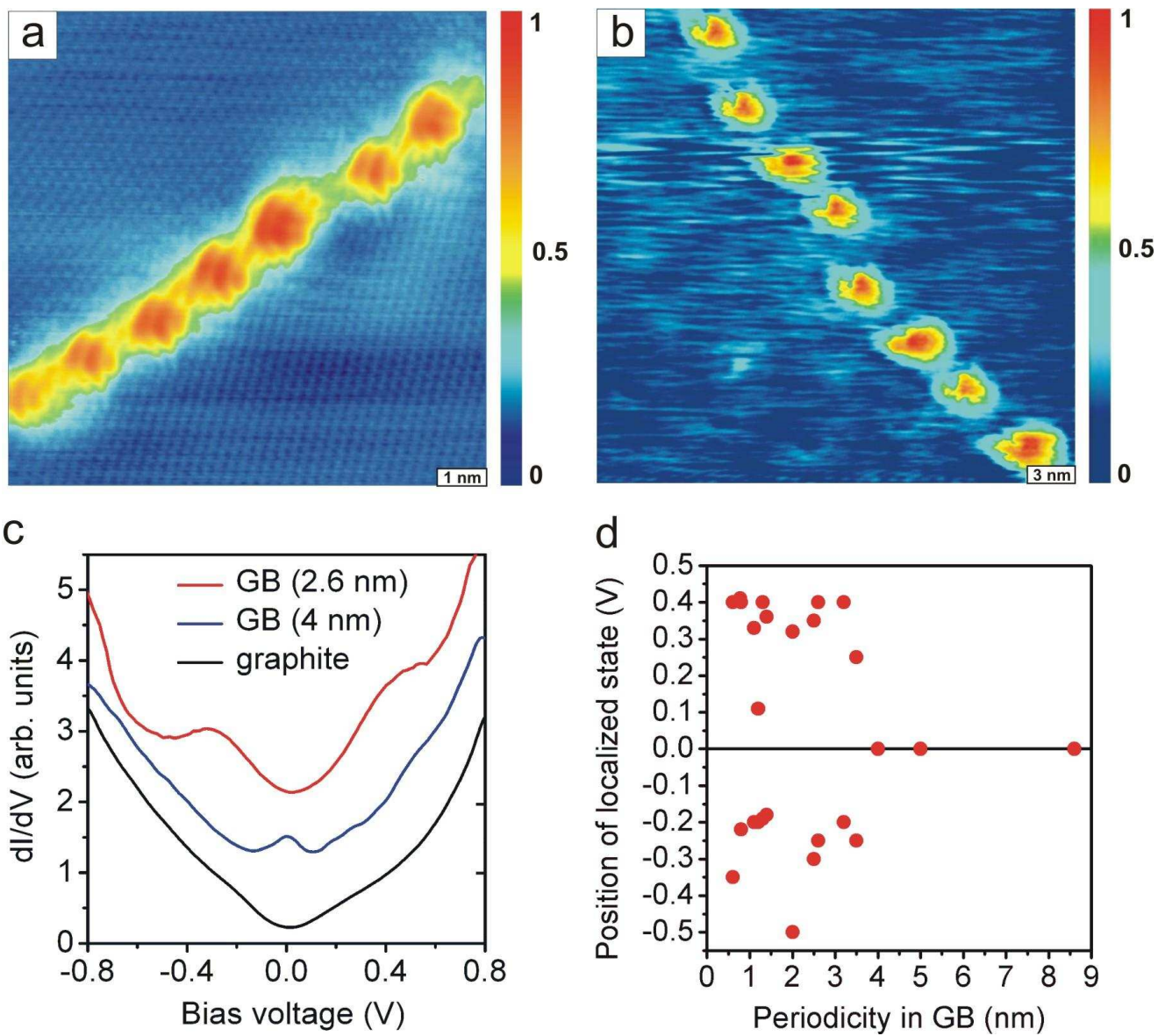

**Figure 3**

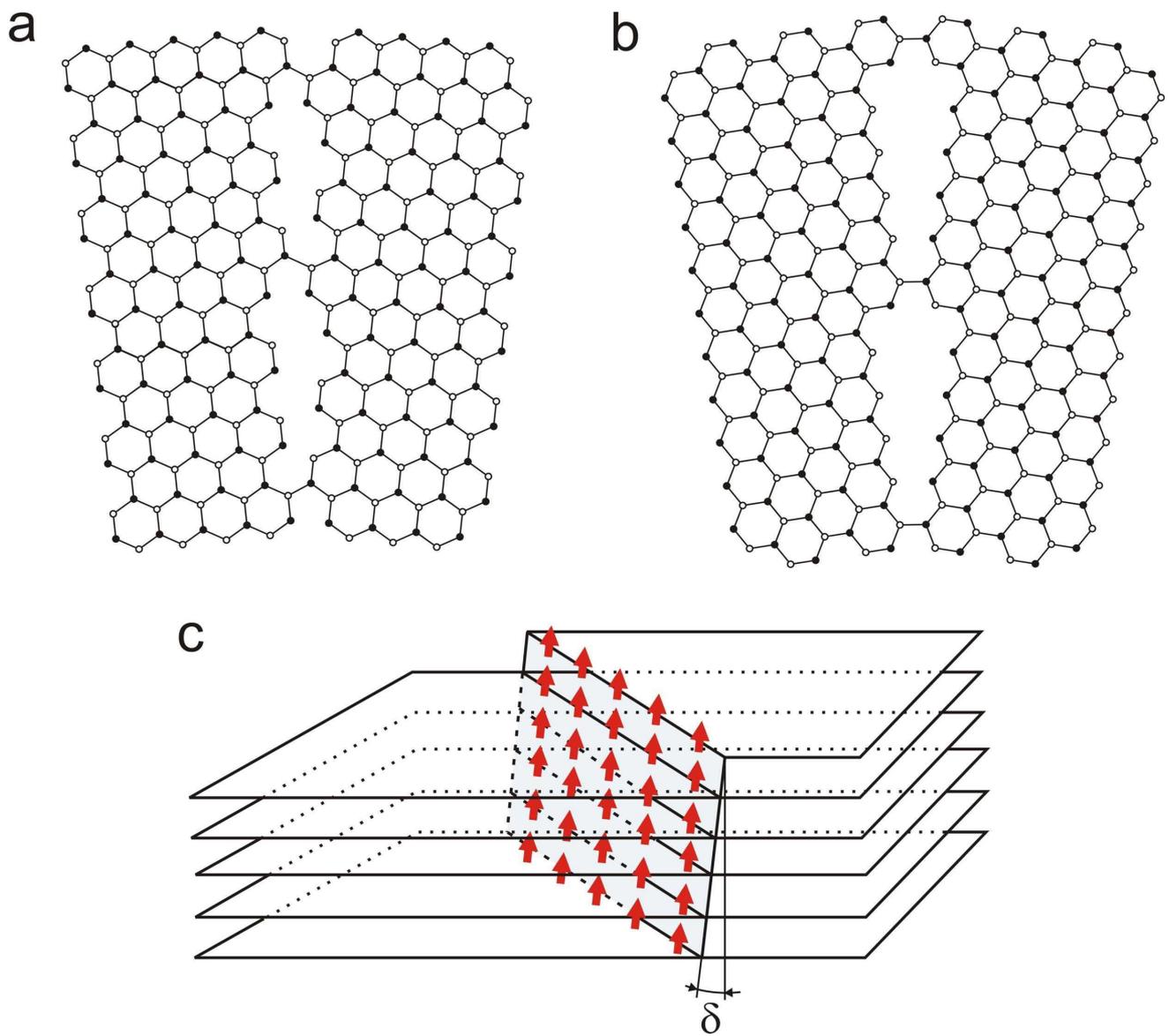

**Figure 4**